\documentclass[%
 reprint,
 amsmath,amssymb,
 aps,
longbibliography
]{revtex4-1}
\usepackage{lipsum}
\usepackage{graphicx}
\usepackage{dcolumn}
\usepackage{bm}
\usepackage{hyperref}
\usepackage{natbib}
\begin{document}

\title{Rydberg-State-Resolved Resonant Energy Transfer in Cold Electric-Field-Controlled Intrabeam Collisions of NH$_3$ with Rydberg He Atoms}

\author{K. Gawlas}
\author{S. D. Hogan}
\affiliation
{Department of Physics and Astronomy, University College London, Gower Street, London WC1E 6BT, United Kingdom}

\date{\today}

\begin{abstract}
The resonant transfer of energy from the inversion sublevels in NH$_3$ to He atoms in triplet Rydberg states with principal quantum number $n=38$ has been controlled using electric fields below 15~V/cm in intrabeam collisions at translational temperatures of $\sim1$~K. The experiments were performed in pulsed supersonic beams of NH$_3$ seeded in He at a ratio of 1:19. The He atoms were prepared in the metastable 1s2s\,$^3$S$_1$ level in a pulsed electric discharge in the trailing part of the beams. The velocity slip between the heavy NH$_3$ and the lighter metastable He was exploited to perform collision studies at center-of-mass collision speeds of $\sim70$~m/s. Resonant energy transfer in the atom-molecule collisions was identified by Rydberg-state-selective electric-field ionization. The experimental data have been compared to a theoretical model of the resonant dipole-dipole interactions between the collision partners based on the impact parameter method.
\end{abstract}

\pacs{}

\maketitle


In the study of gas-phase chemical dynamics at low temperature, i.e., when $T\lesssim1$~K, long-range interparticle interactions, which may be exploited to regulate access to short-range reaction processes, are of particular interest. These interactions can be isotropic or anisotropic, and arise as a result of the static electric dipole polarizabilities or static electric dipole moments of the collision partners, or resonant dipole$-$dipole interactions between them. The large static electric dipole polarizabilities, static electric dipole moments, and electric dipole transition moments associated with high Rydberg states of atoms and molecules have made them ideally suited to the study of long-range interactions of these kinds in ultracold atomic gases. This has led to the synthesis of long-range diatomic molecules with equilibrium internuclear distances of $\sim1~\mu$m, known as Rydberg macro-dimers, which are formed as a result of the interactions between pairs of Rydberg atoms~\cite{Farooqi2003,Overstreet2009,Deiglmayr2014}. In experiments performed at high atom number densities, $\sim10^{11}$~cm$^{-3}$, the properties of the Rydberg states that give rise to strong long-range interactions, i.e., the large spatial extent of the Rydberg electron wave functions and the high density of alternate parity states, have allowed the photoassociation of a class of ultralong-range Rydberg molecules in which an excited Rydberg atom is bound to a ground-state atom located within the Rydberg electron orbit~\cite{Bendkowsky2009,Gaj2014}.

The evolution of studies of long-range interactions of cold Rydberg atoms or molecules to include those with cold molecules in their electronic ground-states allows access to a distinct set of molecular interactions that occur on several different length scales. (i) At long range, where the spatial separation, $R$, between the centers of mass of the ground-state molecule and the atom or molecule in the Rydberg state is much larger than the spatial extent of the Rydberg electron wave function, i.e., $R \gg \langle r \rangle$, where $\langle r \rangle$ is the expectation value of the radial position operator acting on the Rydberg electron wave function, van der Waals and resonant dipole$-$dipole interactions dominate. (ii) On the intermediate length scale, where $R \simeq \langle r \rangle$ and the ground-state particle is located within the Rydberg electron charge distribution, the formation of ultralong-range Rydberg molecules and electron transfer between the collision partners can occur. (iii) At short range, where $R \ll \langle r \rangle$, the most significant interaction is that of the ground-state molecule with the ion core of the Rydberg atom or molecule.

Experimental studies at each of these length scales have been carried out. Resonant energy transfer in collisions between atoms in Rydberg states (Xe, Rb, and He) and polar ground-state molecules (NH$_{3}$, HF, HCl, and CO) emanating from effusive sources operated at 300~K have been studied~\cite{Smith1978,Kellert1980,Petitjean1986,Higgs1981,Stebbings1981,Kal1987,Petitjean1984}. This work included, most recently, observations of electric-field-controlled energy transfer from the inversion sublevels in NH$_{3}$ to Rydberg states in He and Rb~\cite{Zhelyazkova.2017,Zhelyazkova.2017b,Jarisch2018}. The formation of negative ions by electron transfer in collisions of atoms in Rydberg states with polar ground-state molecules, such as acetonitrile and cyclobutanone, at collision energies of $\sim1$~eV have also been investigated~\cite{Desfrancois1994}. Short-range ion$-$molecule reactions, e.g., the H$_{2}^{+}$~+~H$_{2}~\rightarrow$~H$_{3}^{+}$~+~H reaction, have been studied at a translational temperature of $\sim8$~K in intrabeam experiments with supersonic molecular beams containing ground-state and Rydberg H$_{2}$~\cite{Pratt1994}. More recently, this reaction has been performed at collision energies as low as $E_{\mathrm{coll}}/hc=0.24$~cm$^{-1}$ ($E_{\mathrm{coll}}/k_{\mathrm{B}} \simeq300$~mK) in experiments in which beams of Rydberg H$_{2}$ were merged with beams of ground-state H$_{2}$ using the methods of Rydberg-Stark deceleration~\cite{Allmendinger2016}.

\begin{figure*}
\includegraphics[width=0.9\linewidth]{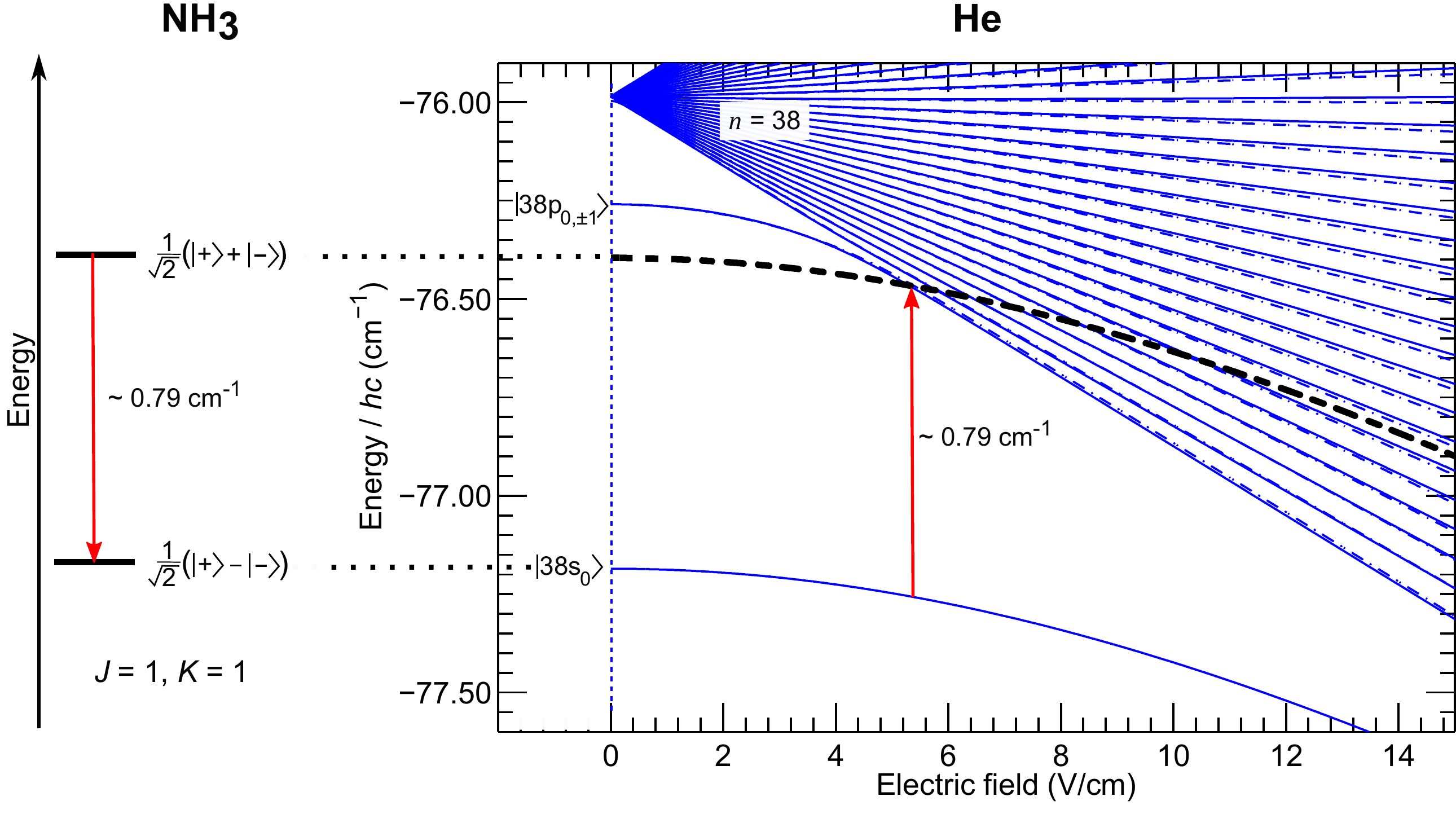}
\caption{Energy-level structure associated with the 0.79 cm$^{-1}$ ground-state inversion transition in NH$_{3}$ (left), and the triplet Rydberg states in He near $n = 38$ (right). In the Stark map on the right sub-levels for which $m_{\ell}$ = 0 ($|m_{\ell}|$ = 1) are indicated by the continuous (dash-dotted) curves. The thick dashed curve 0.79~cm$^{-1}$ above the $\big|$38s$_0^\prime$$\big\rangle$ state, corresponds to the energy of the $\big|$38s$_0^\prime$$\big\rangle$ state offset by the NH$_{3}$ inversion transition wavenumber.} \label{fig1}
\end{figure*}

Here we report for the first time intrabeam collision studies~\cite{Amarasinghe2017,Perreault2017} that take place at translational temperatures of $\sim1$~K, involve Rydberg He atoms and ground-state NH$_{3}$ molecules, and in which fully Rydberg-state resolved resonant energy transfer has been observed. This energy transfer arises as a result of the resonant electric dipole-dipole interaction between the collision partners, where the dipole-dipole interaction energy may be expressed as 
\begin{equation}
V_{\mathrm{dd}} (\vec{R}) = \frac{1}{4\pi\epsilon_0} \Bigg[\frac{\mu_{\mathrm{He}}\mu_{\mathrm{NH_{3}}}}{R^{3}} - \frac{3(\vec{\mu}_{\mathrm{He}}\cdot\vec{R})(\vec{\mu}_{\mathrm{NH_{3}}}\cdot\vec{R})}{R^{5}}\Bigg]
\end{equation}
with $\vec\mu_{\mathrm{He}}$ and $\vec\mu_{\mathrm{NH_{3}}}$ the electric dipole transition moments in He and NH$_{3}$, respectively, $\mu=|\vec{\mu}|$, and $\epsilon_{0}$ the vacuum permittivity. The results reported here represent an important step toward studies in which the long-range electric dipole interactions that lead to resonant energy transfer are exploited to regulate access to the shorter range processes of electron transfer and ion-molecule reactions in this collision system. The observation of the transfer of energy from the internal degrees of freedom in NH$_{3}$ to the Rydberg atoms represents a dissipative process by which the internal degrees of freedom of the molecule may be considered to be cooled. It is therefore of relevance to the realization of  schemes that have been proposed to use Rydberg atoms as a refrigerant to cool~\cite{Huber2012,Zhao2012}, and manipulate the internal degrees of freedom of polar ground-state molecules~\cite{Kuznetsova2011}. The results reported may also be interpreted as a method of non-destructive detection of the ground-state molecules~\cite{Zeppenfeld2017}.

The apparatus used here is similar to that described in Ref.~\cite{gawlas2019}. The experiments were performed in pulsed supersonic beams of NH$_{3}$ seeded in He at a ratio of 1:19. To permit intrabeam collision studies involving the triplet Rydberg states in He, atoms in the metastable 1s2s $^{3}$S$_{1}$ level were first generated in a pulsed electric discharge at the exit of a pulsed valve. This discharge was implemented by applying a pulsed potential of $+220$~V to a metal anode maintained at an offset potential of $+150$~V and located $\sim1$~mm from the valve exit. This was applied for 30~$\mu$s in the trailing part of the $\sim200$-$\mu$s-long gas pulse. The discharge was seeded with electrons generated continuously by a heated tungsten filament $\sim2.5$~cm downstream from the valve. After passing through a 2-mm-diameter skimmer and an electrostatic charged particle filter, the molecular beam entered between two parallel copper electrodes. In the region between these electrodes the metastable He atoms were excited to the 1s38s\,$^{3}$S$_{1}$ level (denoted $\big|$38s$_{0}$$\big\rangle$ in the following, where the subscript indicates the value of $m_{\ell}$) through the 1s3p\,$^{3}$P$_{2}$ intermediate level using a resonance-enhanced two-color two-photon excitation scheme in zero electric field. This scheme was implemented by stabilizing a CW laser to the wavenumber of the 1s2s\,$^{3}$S$_{1}$~$\rightarrow$~1s3p\,$^3$P$_2$ transition at 25$\,$708.587$\,$6~cm$^{-1}$, while a second CW laser subsequently drove the 1s3p\,$^{3}$P$_{2}$~$\rightarrow$~1s38s\,$^3$S$_1$ transition at 12668.919~cm$^{-1}$. The $\big|$38s$_{0}$$\big\rangle$ state (quantum defect $\delta_{38\mathrm{s}}=0.296\,683$~\cite{Drake1999}) was chosen for these experiments because transitions from it to the $\big|$38p$_{0,\pm1}^\prime$$\big\rangle$ state ($\delta_{38\mathrm{p}}=0.068\,347$~\cite{Drake1999}) (the prime indicates the $\ell$-mixed character attained in the presence of an electric field) can be readily tuned into resonance with the ground-state inversion transitions in NH${_3}$ using weak electric fields as shown in Figure~\ref{fig1}. In this particular case the resonance condition is achieved in a field of 5.41~V/cm. After laser photoexcitation, pulsed electric fields of between 0.5 and 14~V/cm were applied to control and study the resonant transfer of energy in collisions between the Rydberg He and the ground-state NH$_{3}$ in the beam. Following an interaction time of 12~$\mu$s, the Rydberg atoms were ionized upon the application of a slowly rising pulsed negative potential to one of the parallel electrodes in the interaction region. The ionized electrons were then accelerated through an aperture in the opposite electrode to a microchannel plate (MCP) detector. This Rydberg-state-selective detection scheme was operated in such a way that individual low-$|m_{\ell}|$ Rydberg states could be identified from their distinct ionization electric fields~\cite{Zhelyazkova.2017}.

\begin{figure}
\includegraphics[width=0.95\linewidth]{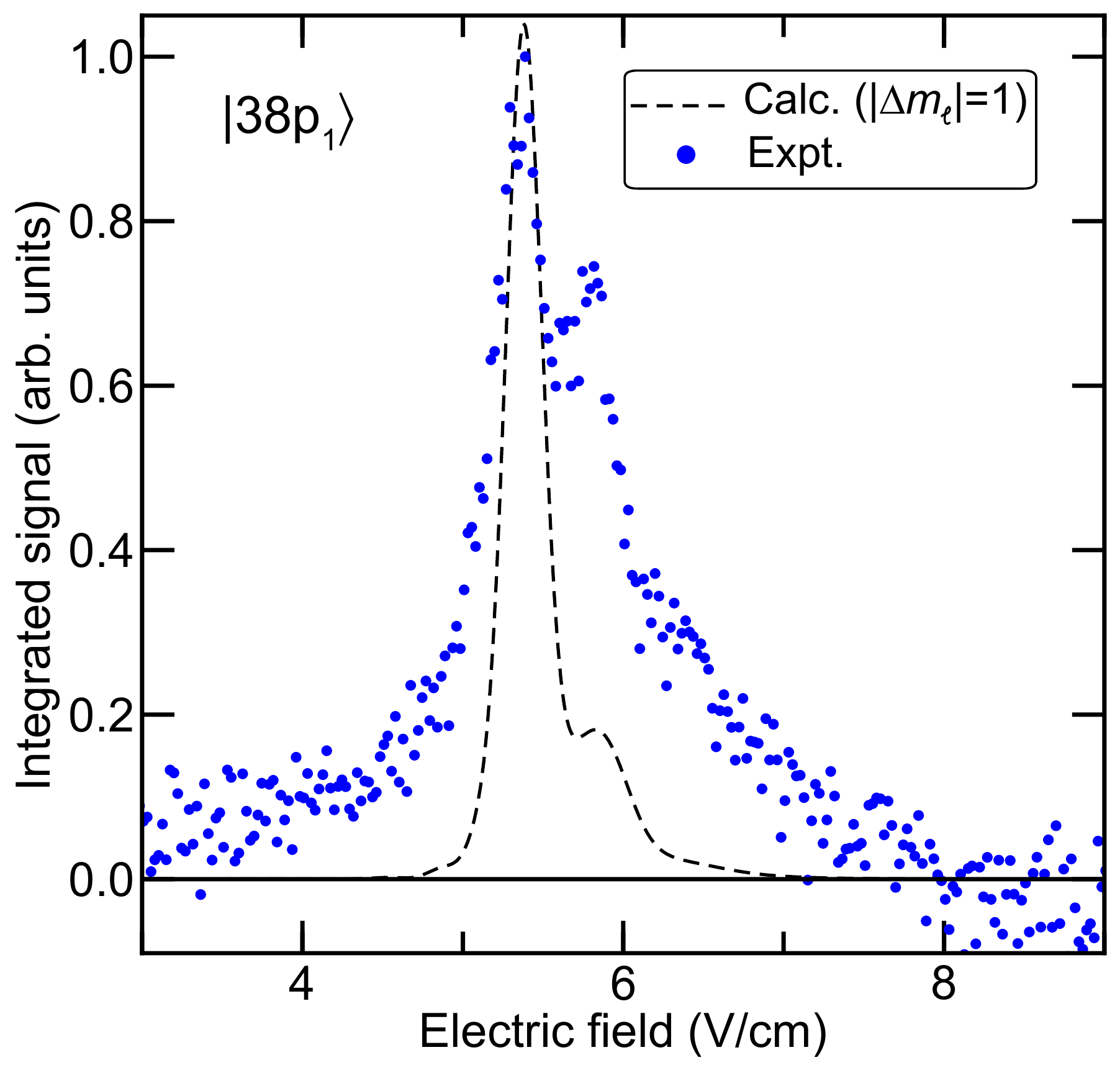}
\caption{Integrated $\big|$38p$_{1}$$\big\rangle$ electron signal (blue points) recorded following collisions in the electric fields indicated on the horizontal axis. The resonance lineshape calculated considering only collision-induced energy transfer from the $\big|$38s$_0^\prime$$\big\rangle$ state to the $\big|$38p$_1^\prime$$\big\rangle$ state is indicated by the dashed curve.} \label{fig2}
\end{figure}

\begin{figure*}
\includegraphics[width=0.7\linewidth]{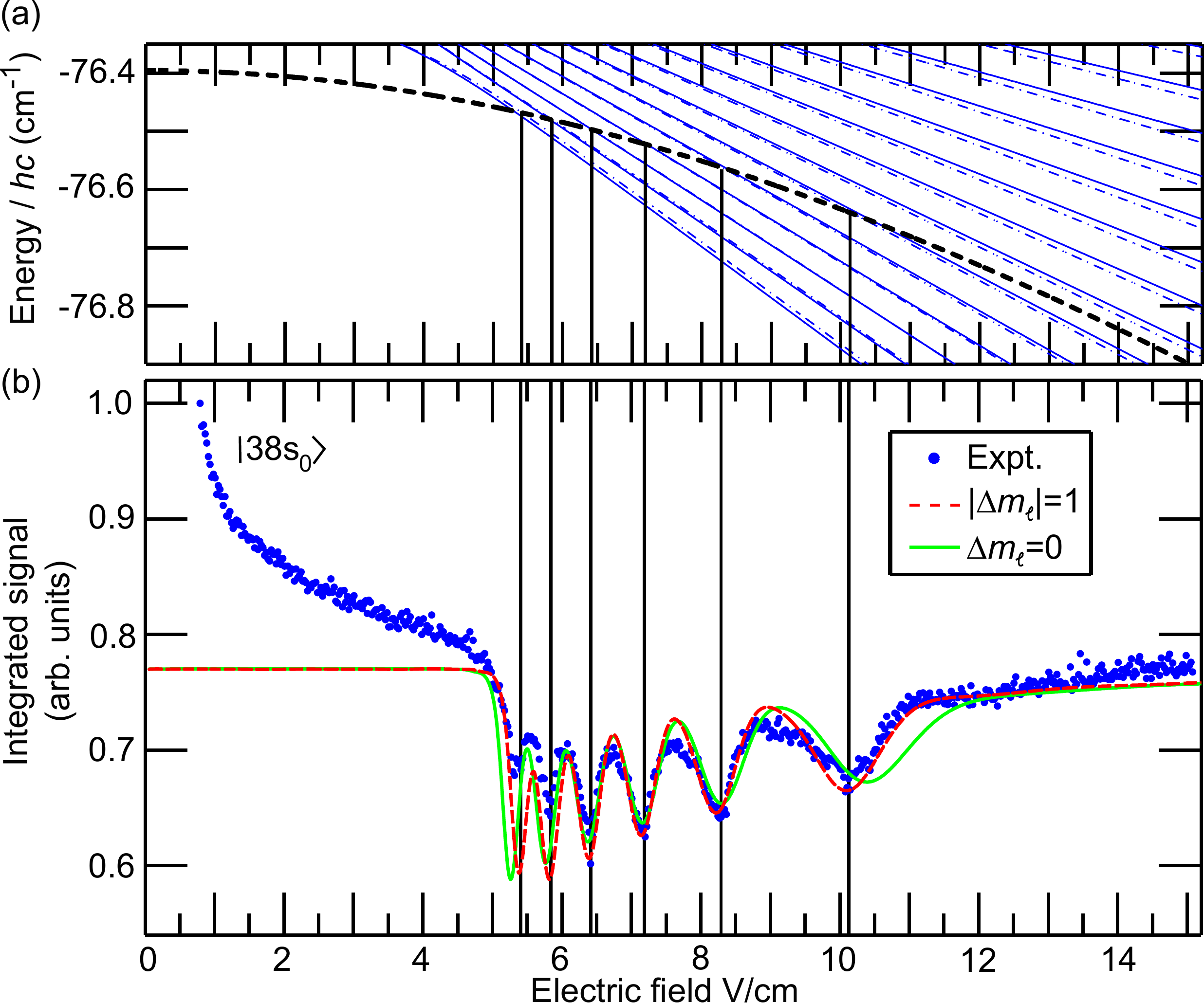}
\caption{(a) A portion of the Stark map in Figure~\ref{fig1}, and (b) the integrated $\big|$38s$_{0}$$\big\rangle$ electron signal (blue points) recorded following collisions in the electric fields indicated on the horizontal axis. The calculated intensities of the resonances that result from transitions to $m_\ell =0$ ($|m_\ell|=1$) states are shown as the continuous (dashed) curves. The vertical lines connecting panels (a) and (b) indicate the electric fields for which transitions from the $\big|$38s$_0^\prime$$\big\rangle$ state in He are resonant with the inversion transitions in NH$_{3}$.} \label{fig3}
\end{figure*}

In zero electric field the electric dipole transition moments associated with the $|38\mathrm{s}_{0}\rangle\rightarrow|38\mathrm{p}_{1}\rangle$ transition in He and the inversion transitions in NH$_{3}$ are $\mu_{ 38\mathrm{s}_{0}-38\mathrm{p}{_1}}=2976$~D and $\mu_{\mathrm{NH_{3}}}=1.4$~D, respectively. These transitions are not resonant in zero field as can be seen in Figure~\ref{fig1}. The integrated $|38\mathrm{p}_1\rangle$ electron signal recorded in the experiments following collisions in electric fields between 3 and 9~V/cm is shown as the points in Figure~\ref{fig2}. The increase in the population of the $\big|$38p$_{1}$$\big\rangle$ state in fields between 4 and 7~V/cm, with a maximum at 5.39~V/cm, results from the resonant transfer of energy from the inversion sublevels in NH$_{3}$ to the He atoms. From the Stark map in Figure~\ref{fig1}, it can be seen that the $\big|$38s$_0^\prime$$\big\rangle \rightarrow$ $\big|$38p$_1^\prime$$\big\rangle$ transition is resonant with the inversion transitions in a field of 5.41~V/cm. This lies within 0.02~V/cm of the observed resonance in Figure~\ref{fig2}. In this field $\mu_{ 38\mathrm{s}_0^\prime- 38\mathrm{p}_1^\prime}\simeq1370$~D and $\mu_{\mathrm{NH_{3}}}$ is unchanged from its zero-field value. The measured electric-field dependence of the energy transfer process has been compared to the results of calculations based on the interaction of two electric dipoles corresponding to the He and NH$_{3}$ transition dipole moments. These calculations were performed, as described in detail in Refs.~\cite{gallagher1992,Zhelyazkova.2017b}, using the impact parameter method~\cite{Seaton1962}. The experimental data are in good agreement with the results of these calculations (dashed curve in Figure~\ref{fig2}). The broadening of the resonance on the high field side in the calculated curve is a consequence of the rotational state dependence of the inversion intervals. The broadening observed in the experimental data arises from a combination of this, the limited Rydberg-Stark-state resolution in the detection of the $\big|$38p$_{1}$$\big\rangle$ atoms, and contributions from energy-level shifts arising from the dipole-dipole interactions between the atoms and molecules that are not accounted for in the calculations.

The resonant increase in the $\big|$38p$_{1}$$\big\rangle$ electron signal seen in Figure~\ref{fig2} corresponds to a decrease in the electron signal from the $\big|$38s$_{0}$$\big\rangle$ state over the same range of electric fields. This can be seen in the data presented in Figure~\ref{fig3}(b). It is evident from these data that the transition to the $\big|$38p$_1^\prime$$\big\rangle$ state does not represent the only resonance that occurs in the experiments. Five additional resonances, in fields between 5 and 11.5~V/cm, are seen. These correspond to transitions from the $\big|$38s$_0^\prime$$\big\rangle$ state to consecutive $\ell$-mixed Rydberg-Stark states that lie above the $\big|$38p$_1^\prime$$\big\rangle$ state in energy. The spectral overlap of these transitions with the NH$_{3}$ inversion transitions can be seen by following the thick dashed curves in Figure~\ref{fig1} and Figure~\ref{fig3}(a). The fields in which the transitions to the $\big|$38p$_1^\prime$$\big\rangle$ state, and the five adjacent Rydberg-Stark states, coincide with the inversion transition wavenumber are indicated in Figure~\ref{fig3}(a) by the vertical lines. It can be seen from the positions of these lines in Figure~\ref{fig3}(b) -- which is presented with the same horizontal scale -- that these crossing points correspond exactly to electric fields at which the resonances are observed in the experimental data. As the electric field is increased the observed resonances broaden because the induced electric dipole moment associated with the Stark shift of the $\big|$38s$_0^\prime$$\big\rangle$ state tends toward those of the Stark states to which resonant energy transfer occurs. Consequently, the wavenumber interval between the $\big|$38s$_0^\prime$$\big\rangle$ state and these Stark states changes more slowly with field strength. 

Calculated spectra for $\Delta m_{\ell}=0$ ($\Delta m_{\ell}=\pm1$) transitions between the Rydberg states are shown as the continuous (dashed) curves in Figure~\ref{fig3}(b). In the calculations, the features observed at low electric field strengths, i.e., in fields between 5 and 7~V/cm, are most sensitive to the center-of-mass collision speed. Best agreement between the experimental and calculated data was found for collision speeds of $\sim70$~m/s, which corresponds to a typical center-of-mass collision energy $E_{\mathrm{coll}}/k_{\mathrm{B}} \simeq1$~K. The widths of the resonances in the data in Figure~\ref{fig3}(b) also depend on the NH$_{3}$ rotational temperature. The resonances in fields above 7.5~V/cm are most sensitive to this temperature, with the broad curve in the data between fields of 11 and 14~V/cm especially so. It was found that best agreement between the experimental and calculated data, particularly in this higher electric field region, occurred for a rotational temperature of $\sim50$~K. In seeded supersonic molecular beams of the kind used in these experiments, rotational temperatures of $\sim1$~K are typical. The NH$_3$ rotational temperature of 50~K, inferred by comparing the experimental data with the results of calculations, is attributed to collision-induced heating in the region close to the valve. This arises from a combination of collisions involving the NH$_3$ molecules and the electrons generated to seed the discharge, which were continually accelerated through this region by the $+150$~V offset potential applied to the anode, and turbulence caused by the presence of the mechanical components required for the operation of the discharge. It is expected that future refinements to the operation of the discharge source will allow lower rotational temperatures to be achieved. In Figure~\ref{fig3}(b), the results of the calculations performed for $\Delta m_{\ell}=\pm1$ transitions between the Rydberg states is in slightly better agreement with the experimental data than those for the $\Delta m_{\ell}=0$ transitions. This is most notable in electric fields above 8~V/cm, and may be indicative of a preferred orientation of the NH$_{3}$ dipole moments in the interaction region in the experiments.

In general, the widths and intensities of the measured and calculated resonances in Figure~\ref{fig3}(b) are in good quantitative agreement with the exception of the intensities of the first two features at 5.4 and 5.8~V/cm. These are weaker in the experiments than in the calculations. We attribute this discrepancy to a non-thermal NH$_3$ rotational state population. The calculated curves in Figure~\ref{fig3}(b) do not account for the notable decrease in the $\big|$38s$_{0}$$\big\rangle$ electron signal seen in fields between 0.5 and 5~V/cm. This suggests that additional transitions, other than those associated with NH$_{3}$ inversion, also transfer population out of the $\big|$38s$_0^\prime$$\big\rangle$ state in the experiments. This is supported by the presence of additional features in the electric field ionization profiles of the Rydberg states (not shown) that correspond to the ionization of atoms in higher $n$ states (e.g., $n=46-51$). Transitions between excited rotational states in NH$_3$ which are resonant with $n$-changing transitions in He can also give rise to energy transfer~\cite{Smith1978} and appear to be responsible for this overall decrease in the $\big|$38s$_{0}$$\big\rangle$ electron signal observed in low fields.

The spectral widths of the sharp resonances in Figure~\ref{fig3}(b) can be determined from the calculated Stark shifts of each transition. The full-width-at-half-maximum of these range from 0.015~cm$^{-1}$ ($\equiv460$~MHz), for the resonance associated with the transition to the $\big|$38p$_1^\prime$$\big\rangle$ state at 5.39~V/cm, to 0.021~cm$^{-1}$ ($\equiv640$~MHz) for the transition observed in the highest field. The impact parameter method used in the calculations in Figure~\ref{fig3}(b) is based on the assumption that energy-level shifts arising from the resonant dipole-dipole interactions between the atoms and the molecules can be neglected. This is a valid assumption at high center-of-mass collision speeds, but less so at low speed where these energy-level shifts are on the same scale as the collision energy. To estimate the significance of these effects in the present experiments, the observed resonance widths may be compared to the typical resonant dipole$-$dipole interaction energy for an NH$_{3}$ molecule located at the edge of the Rydberg electron charge distribution, i.e., at $\langle r \rangle$. This is approximately the closest point of approach between the collision partners after which the NH$_{3}$ molecule enters the Rydberg electron charge distribution and the form of the interaction potential changes. Because $\langle r_{\mathrm{38s}} \rangle \simeq$~110~nm, and $\mu_{ 38\mathrm{s}_0^\prime- 38\mathrm{p}_1^\prime}(F=5.39~\mathrm{V/cm})\simeq1370$~D, the corresponding dipole$-$dipole energy shift is $\Delta E_{\mathrm{dd}}/hc\simeq0.007$~cm$^{-1}$ ($\equiv220$~MHz). This energy shift is on the same order of magnitude as the widths of the resonances in the experiments. Therefore, the conditions under which the experiments were performed must represent the limit of the range of validity of the impact parameter method used in the analysis of the data which does not account for effects of these types of energy-level shifts. This observation provides a strong motivation for a more complete theoretical treatment of the dipole$-$dipole interactions in this collision system. Such work will be essential for the accurate interpretation of future experiments at lower collision energies.

In conclusion, we have observed Rydberg-state-resolved and electric-field-controlled resonant energy transfer in collisions of ground-state NH$_{3}$ molecules with Rydberg He atoms at translational temperatures of $\sim1$~K. The experiments were performed in a single pulsed supersonic beam of NH$_{3}$ seeded in He. The measured data exhibit resonance widths of  $\sim0.017$~cm$^{-1}$ ($\equiv500$~MHz). This work paves the way for experiments at lower collision energies in which the resonant dipole$-$dipole interactions between the atoms and the molecules are more strongly affected by the energy-level shifts of the collision partners and may be exploited to regulate access to short-range chemical dynamics. It is foreseen that in future experiments the center-of-mass collision energies in these experiments may be manipulated using the methods of Rydberg-Stark deceleration~\cite{Hogan2016}.

\begin{acknowledgements}
This work is supported by the European Research Council (ERC) under the European Union's Horizon 2020 research and innovation program (Grant No. 683341).
\end{acknowledgements}


\bibliography{references}

\end{document}